\documentclass[10pt,aps,pra,twocolumn,groupedaddress,longbibliography,reprint]{revtex4-1}

\usepackage[english]{babel}
\usepackage[utf8]{inputenc}
\usepackage[colorlinks=true,citecolor=blue]{hyperref}

\usepackage{amsmath,amssymb,mathtools,bbm}

\usepackage{graphicx}
\graphicspath{{./figures/}}
\usepackage[shortlabels,inline]{enumitem}

\DeclarePairedDelimiter{\abs}{\lvert}{\rvert}
\DeclarePairedDelimiter{\bra}{\langle}{\rvert}
\DeclarePairedDelimiter{\ket}{\lvert}{\rangle}
\DeclarePairedDelimiterX{\braket}[2]{\langle}{\rangle}{#1 \delimsize\vert #2}
\DeclarePairedDelimiterX{\comm}[2]{[}{]}{#1, #2}
\DeclareMathOperator{\tr}{tr}
\newcommand{\Tr}[1]{\tr\mathopen{}\bigl[ #1 \bigr]\mathclose{}}

\newcommand{\nsum}{\sum_{\mathclap{n=1}}^N}
\newcommand{\ksum}{\sum_{\mathclap{k=1}}^n}
\newcommand{\nmsum}{\sum_{\mathclap{n,m=1}}^N}
\newcommand{\period}{\tau}
\newcommand{\ergo}{\mathcal E}
\newcommand{\ergoflux}{\mathcal J}
\newcommand{\flux}{\mathcal I}

\newcommand{\idt}{\,dt}

\begin{document}

\title{Thermodynamics of Cyclic Quantum Amplifiers}
\author{Paul Menczel}
\author{Christian Flindt}
\author{Kay Brandner}
\affiliation{Department of Applied Physics, Aalto University, 00076 Aalto, Finland}

\begin{abstract}
We develop a generic model for a cyclic quantum heat engine that makes it possible to coherently amplify a periodically modulated input signal without the need to couple the working medium to multiple reservoirs at the same time.
Instead, we suggest an operation principle that is based on the spontaneous creation of population inversion in incomplete relaxation processes induced by periodic temperature variations.
Focusing on Lindblad dynamics and systems with equally spaced energy levels, e.g.~qubits or quantum harmonic oscillators, we derive a general working criterion for such cyclic quantum amplifiers.
This criterion defines a class of candidates for suitable working media and applies to arbitrary control protocols.
For the minimal case of a cyclic three-level amplifier, we show that our criterion is tight and explore the conditions for optimal performance.
\end{abstract}

\maketitle

\section{Introduction}

\begin{figure}
	\centering
	\includegraphics[scale=1]{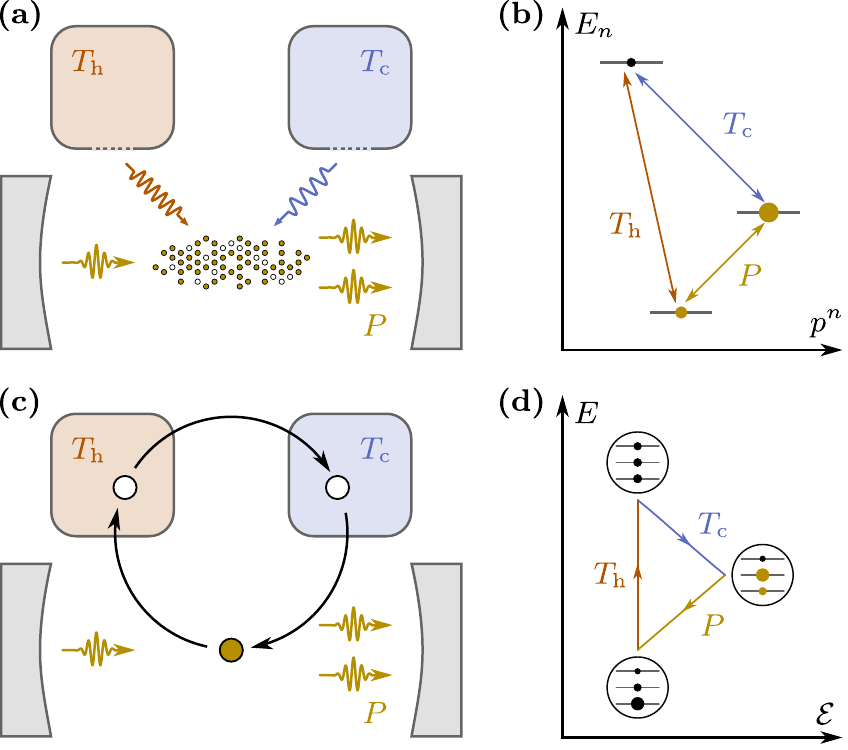}
	\caption[x]{
		Quantum amplifiers.
		\begin{enumerate*}[(a)]
		\item Continuous device using a medium of three-level atoms, whose transitions are selectively coupled to a hot ($T_h$) and cold ($T_c$) reservoir via ideal energy filters;
			the coherent power $P$ is extracted by applying an input signal on resonance with the inverted transition.
		\item Steady-\mbox{state} populations~$p^n$ of the atomic energy levels~$E_n$.
		\item Cyclic device operating in three strokes.
			The internal energy $E$ of a multi-level atom is first increased by injecting heat from a hot reservoir and then reduced in a cold environment to create a population-inverted state with finite \emph{ergotropy} $\mathcal{E}$, see Eq.~\eqref{eq:ThDyn_Ergotropy}.
			In the third stroke, a resonant pulse extracts the power $P$, whereby the system returns to its initial state.
		\item Energy-ergotropy diagram of the amplifier cycle.
		\end{enumerate*}
		Insets show the state of a three-level atom at the beginning of each stroke.
	}
	\label{fig:1}
\end{figure}

Quantum amplifiers generate coherent electromagnetic energy using the stimulated emission of photons in a population-inverted medium~\cite{GevaPhysRevE1994, BoukobzaPhysRevA2006, ClerkRevModPhys2010}.
Early on, Scovil and Schulz-DuBois realized that, when driven by a thermal gradient, such devices can be understood as quantum-mechanical heat engines, whose efficiency is subject to 
the Carnot bound~\cite{ScovilPhysRevLett1959}.
In their approach, the working medium is a collection of three-level atoms, whose transitions are coupled either to a hot or a cold reservoir acting as a source of energy and a sink of entropy, respectively.
A resonant driving field plays the role of a moving piston enabling the extraction of usable work in form of coherent radiation, see Fig.~\ref{fig:1}a--b.

Owing to its universal and transparent structure, this model has contributed significantly to our basic knowledge of energy conversion in the quantum regime~\cite{ScovilPhysRevLett1959, GevaPhysRevE1994, GevaJChemPhys1996, BoukobzaPhysRevA2006, BoukobzaPhysRevLett2007, ClerkRevModPhys2010, SandnerPhysRevLett2012, UzdinPhysRevX2015, UzdinPhysRevAppl2016, LiPhysRevA2017}.
At the same time, it has become a prototype for practical devices like photocells~\cite{ScullyPhysRevLett2010, DorfmanPNAS2013, XuNewJPhys2016, SuPhysRevE2016} and small-scale 
refrigerators~\cite{KosloffJournalofAppliedPhysics2000, PalaoPhysRevE2001, LindenPhysRevLett2010, CleurenPhysRevLett2012}.
Moreover, the three-level amplifier has served as a template for new types of thermal machines that utilize complex quantum effects such as lasing without inversion~\cite{ScullyPhysRevLett2002, ScullyScience2003}, noise-induced coherence~\mbox{\cite{ScullyPNAS2011, HarbolaEPL2012, DorfmanPhysRevE2018}} or electromagnetically induced transparency~\cite{HarrisPhysRevA2016};
	recent proposals include even two-level variants that operate without population inversion using thermal evaporation~\cite{GerasimovPhysRevLett2006}, squeezed driving fields~\cite{GhoshPNAS2017} or two-photon transitions~\cite{GhoshPNAS2018}.
These developments have led to profound theoretical insights over the last years.
They might soon also be tested in practice as coherence-based heat engines can now be realized experimentally~\cite{ZouPhysRevLett2017, KlatzowPhysRevLett2019}.

The ideas of Scovil and Schulz-DuBois have shaped our perception of thermal quantum amplifiers as a distinct sort of heat engines, which operate in a steady state and use reversible energy filters to maintain a population-inverted working medium~\cite{KosloffAnnuRevPhysChem2014, HumphreyPhysicaE2005}.
In this article, we investigate an alternative strategy for coherent power generation:
	we develop and analyze a generic model for a cyclic quantum amplifier.
Resembling a reciprocating heat engine, our device operates in a thermodynamic cycle~\cite{QuanPhysRevE2005, HumphreyPhysicaE2005, QuanPhysRevE2007, VinjanampathyContempPhys2016}, where heat is transferred periodically from a hot to a cold reservoir to create population inversion, see Fig.~\ref{fig:1}c--d.
In contrast to earlier proposals, this working principle does not rely on energy filters.
Instead, it requires at least one metastable energy level, which can be temporarily overpopulated while the system returns to equilibrium.
To capture this condition quantitatively, we derive a general working criterion for cyclic quantum amplifiers, which makes it possible to characterize the applicable working systems without reference to a specific control protocol.

Our manuscript is organized as follows.
In Sec.~\ref{sec:setup}, we introduce the theoretical framework to describe cyclic quantum amplifiers.
In Sec.~\ref{sec:working_crit}, we present our working criterion and explain its physical content.
Mathematical derivations are provided in Sec.~\ref{subsec:derivation}.
In Sec.~\ref{sec:example}, we apply our working criterion to a three-level quantum amplifier and analyze its power and efficiency.
Finally, we conclude in Sec.~\ref{sec:outlook} by discussing future perspectives.

\section{Setup} \label{sec:setup}

\subsection{Cyclic Quantum Amplifiers}

Our setup consists of two basic components:
	a working medium with tunable Hamiltonian
\begin{equation} \label{eq:ThDyn_Hamiltonian}
	H_t \equiv \nsum E_n \ket{n_t} \bra{n_t}
\end{equation}
and a heat source to control the temperature $T_t$ of the environment.
The time-dependence of the energy eigenstates $\ket{n_t}$ is determined by the input signal, while the energy levels $E_n$ are fixed.
This condition ensures that the device exchanges only coherent power with the driving field~\cite{BrandnerPhysRevLett2017}.
At the same time, it reduces the accessible energy content of the system to the maximum amount of work that can be extracted through unitary operations.
This quantity is given by the \emph{ergotropy}~\cite{AllahverdyanEPL2004, Perarnau-LlobetPhysRevX2015, UzdinPhysRevX2018, GhoshEurPhysJSpecTop2019}
\begin{equation} \label{eq:ThDyn_Ergotropy}
	\ergo_t	\equiv \Tr{\rho_t H_t} - \min\nolimits_U \Tr{\rho_t U H_t U^\dagger}
			\equiv E_t - E^{\text{res}}_t
			\geq 0 ,
\end{equation}
where $\rho_t$ denotes the state of the system and $E_t$ its total internal energy.
The residual energy $E^{\text{res}}_t$ is found by evaluating the minimum over all unitary operators $U$.

Taking the time derivative of Eq.~\eqref{eq:ThDyn_Ergotropy} yields
\begin{equation} \label{eq:ThDyn_ErgotropyBalance}
	\dot{\ergo}_t = \ergoflux_t - P_t .
\end{equation}
This balance equation plays the role of the first law of thermodynamics for quantum amplifiers. 
The quantities
\begin{align} \label{eq:ThDyn_ErgotropyPower}
	P_t			&\equiv -\Tr{\rho_t \dot H_t}
				= \nsum \bra{\dot n_t} \comm{H_t}{\rho_t} \ket{n_t}
				\quad\text{and} \\
	\ergoflux_t	&\equiv \Tr{\dot\rho_t H_t} - \dot E^{\text{res}}_t
    			= \nsum \bigl( \bra{n_t} \dot\rho_t \ket{n_t} - \bra{r^n_t} \dot\rho_t \ket{r^n_t} \bigr) E_n \nonumber
\end{align}
correspond to the instantaneous power output and the rate of reservoir-induced ergotropy production~\cite{AlickiJPhysA1979, BrandnerPhysRevLett2017}. 
Here, we have used the ordered spectral decomposition
\begin{equation}
	\rho_t = \nsum r^n_t \ket{r^n_t} \bra{r^n_t}
\end{equation}
of the state $\rho_t$ to evaluate the time derivative of the residual energy.
The energy levels $E_n$ are thereby arranged in ascending order, i.e., $r^n_t \geq r^m_t$ and $E_n \leq E_m$ for $m > n$.
The derivative of the residual energy in Eq.~\eqref{eq:ThDyn_ErgotropyPower} has been evaluated with help of the Hellmann-Feynman theorem \cite{Note1}.
The second expression for $P_t$ follows from Eq.~\eqref{eq:ThDyn_Hamiltonian} and vanishes if the system is in a quasi-classical state, i.e., if $\rho_t$ commutes with $H_t$.
This observation shows that coherent power generation is a genuine quantum phenomenon, which requires the creation of superpositions between the energy levels of the medium~\cite{BrandnerPhysRevLett2017}.

Once the system has settled to a cyclic state, the ergotropy $\ergo_t$ becomes a periodic function of time.
Thus, upon averaging Eq.~\eqref{eq:ThDyn_ErgotropyBalance} over one period $\period$, the mean extracted work becomes
\begin{equation} \label{eq:ThDyn_PowerErgotropy}
	W = \int_0^\period P_t \idt = \int_0^\period \ergoflux_t \idt . 
\end{equation}
This relation shows that a cyclic quantum amplifier can only deliver finite output if the thermal ergotropy production $\ergoflux_t$ becomes positive during its operation cycle.
Hence, it must be possible to drive the system into a population-inverted state by changing the temperature of its environment.
In the following, we will further examine the necessary conditions for this effect.

\subsection{Quantum Ladders}

We now specify the working medium as a quantum ladder with equally spaced energy levels, i.e., we set $E_n = \hbar\omega n$, where $\hbar\omega$ denotes the overall energy scale~\cite{AbergPhysRevLett2014}.
Such systems include, for example, qubits and quantum harmonic oscillators.
In order to describe the interaction of the medium with its environment, we use the well-established Lindblad approach \cite{Scully1997, Breuer2002, KosloffEntropy2013, PekolaNatPhys2015}, which relies on the assumption that the coupling between system and environment is weak and that the driving is slow compared to both the unitary dynamics of the bare system and the relaxation dynamics of the reservoir.
Under these conditions, the time evolution of the state $\rho_t$ is governed by a Markovian quantum master equation~\cite{AlickiJPhysA1979, AlbashNewJPhys2012},
\begin{align} \label{eq:MD_MasterEquation}
	\dot{\rho_t} = -\frac{i}{\hbar} \comm{H_t}{\rho_t}
		&+ \gamma \nu_t \bigl( \comm{L_t \rho_t}{L^\dagger_t} + \comm{L_t}{\rho_t L^\dagger_t} \bigr) / 2 \\
	    &+ \gamma (\nu_t+1) \bigl( \comm{L^\dagger_t \rho_t}{L_t} + \comm{L^\dagger_t}{\rho_t L_t} \bigr) / 2 . \nonumber
\end{align}
Here, the jump operators 
\begin{equation} \label{eq:MD_JumpOperators}
	L_t \equiv \sum_{\mathclap{n=1}}^{\mathclap{N-1}} \ell_n \ket{(n{+}1)_t} \bra{n_t} 
	\quad\text{and}\quad
	L_t^\dagger \equiv \sum_{\mathclap{n=1}}^{\mathclap{N-1}} \ell_n \ket{n_t} \bra{(n{+}1)_t} 
\end{equation}
describe the exchange of photons between system and reservoir, assuming for the sake of simplicity that the weighting factors $\ell_n$ are real.
The rate $\gamma > 0$ determines the average frequency of emission and absorption events and the Bose-Einstein factors $\nu_t \equiv 1 / (e^{\hbar\omega / k_B T_t}-1)$ ensure thermodynamic consistency~\cite{BrandnerPhysRevE2016}.
For a quantitative review of the conditions for the validity of the master equation \eqref{eq:MD_MasterEquation}, see e.g.\ the supplemental material of \cite{BrandnerPhysRevLett2020} and the references therein.

\section{Working Criterion}

\subsection{Bound on Ergotropy Production} \label{sec:working_crit}

The operation principle of our engine relies on the possibility to thermally create population inversion in the system.
In order to achieve this effect, the working medium has to satisfy a minimal condition that follows from the upper bound on the reservoir-induced ergotropy production
\begin{equation} \label{eq:main_bound}
	\ergoflux_t \leq \hbar\omega\gamma \nsum \bigl( r^n_t-r^{n+1}_t \bigr) \Phi_n ,
\end{equation}
which we derive in the next section.
Here, we have used the definition $r^{N+1}_t \equiv 0$ and introduced the system specific constants
\begin{equation} \label{eq:phi_n}
	\Phi_n \equiv \ksum \Bigl( (k-n) \pi_k\bigl( \bigl\{ \ell_{m-1}^2{-}\ell_m^2 \bigr\} \bigr) - \pi_k\bigl( \bigl\{ \ell_{m-1}^2 \bigr\} \bigr) \Bigr) ,
\end{equation}
where the function $\pi_k(M)$ returns the $k^{\text{th}}$-lowest element of the set $M$, the index $m$ assumes values $1\!\leq\!m\!\leq N$, and $\ell_0 \equiv \ell_N \equiv 0$.
Since, by assumption, $r^n_t \geq r^{n+1}_t$, it follows that $\ergoflux_t$ can become positive only if 
\begin{equation}\label{eq:NPC_BoundPhiMax}
	\Phi_{\text{max}} \equiv \max\nolimits_n \Phi_n > 0.
\end{equation}
If $\Phi_{\text{max}}$ is zero or negative, $\ergoflux_t$ cannot be positive at any time during the cycle and it follows from Eq.~\eqref{eq:ThDyn_PowerErgotropy} that work extraction is impossible.
The criterion~\eqref{eq:NPC_BoundPhiMax} thus provides a necessary condition that makes it possible to identify suitable working media for cyclic quantum amplifiers.
Quite remarkably, it depends neither on the Hamiltonian $H_t$, the temperature profile $T_t$, nor on the specific state $\rho_t$.
Hence, it can be used to determine whether or not relaxation-induced population inversion can occur in a given system.

Two natural choices of potential working media are qubits and quantum harmonic oscillators.
Applying our criterion to a qubit, i.e., $N = 2$, we find
\begin{equation}
	\Phi_1^{\text{QB}} = \Phi_2^{\text{QB}} = \Phi_{\text{max}}^{\text{QB}} = 0 .
\end{equation}
For the harmonic oscillator, which features infinitely many energy levels and weighting factors $\ell_n = \sqrt{n}$, we similarly obtain
\begin{equation}
	\Phi_n^{\text{HO}} = \ksum (n-2k+1) = \Phi_{\text{max}}^{\text{HO}} = 0 .
\end{equation}
Hence, our criterion rules out both qubits and harmonic oscillators as working substances of cyclic quantum amplifiers.

\begin{figure*}
	\centering
	\includegraphics[scale=1]{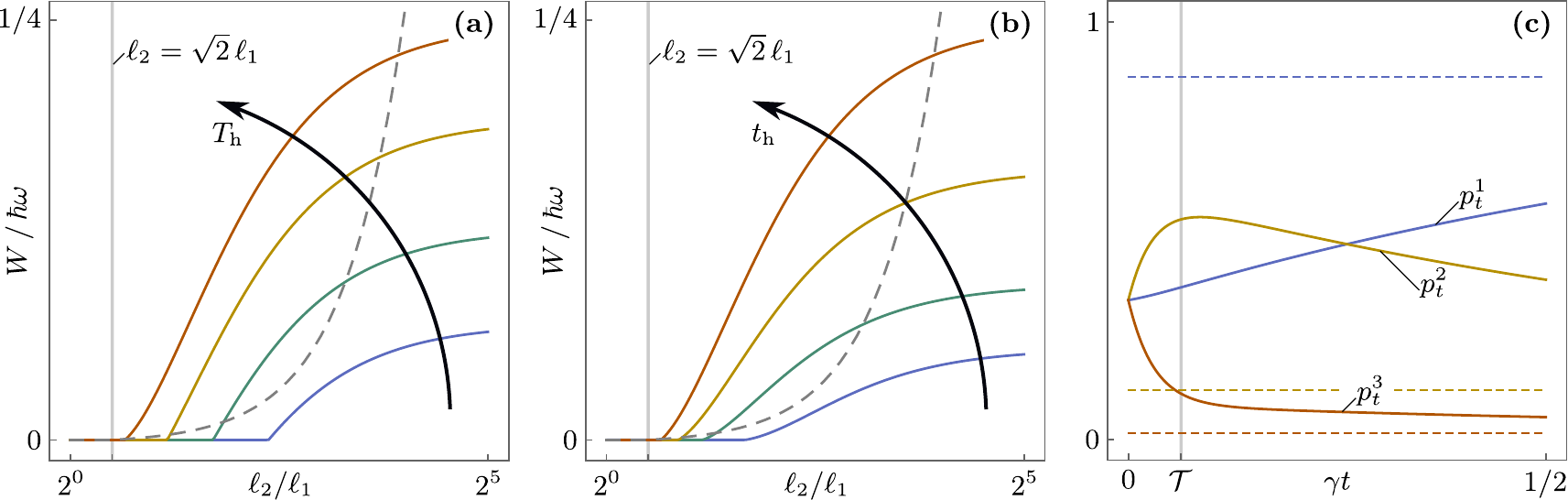}
	\caption[x]{Work output of a cyclic quantum amplifier using a three-level quantum ladder and the operation protocol described in the main text.
		\begin{enumerate*}[(a)]
		\item Work output $W$ per cycle as a function of the ratio $\ell_2 / \ell_1$ of jump weights for fixed input time $t_{\text{h}} = 100\, \gamma^{-1}$ and four different values $k_B T_{\text{h}} = 100, 10, 5, 3.5\, \hbar\omega$ of the hot temperature.
			The thick arrow indicates the direction in which $T_{\text{h}}$ increases.
			The dashed gray line shows the quantity $\Phi_{\text{max}} \cdot 10^{-3}$, which is zero for $\ell_2 \leq \sqrt{2} \ell_1$.
		\item Same plot as in (a) with $k_B T_{\text{h}} = 100\, \hbar\omega$ fixed and $t_{\text{h}} = 1, 0.01, 0.005, 0.002\, \gamma^{-1}$.
		\item Time evolution of the level populations during the conversion stroke 2 for $\ell_2 / \ell_1 = 5$, in the limit of large hot temperature $T_{\text{h}}$ and input time $t_{\text{h}}$, i.e., $T_{\text{h}}, t_{\text{h}} \to \infty$.
			A spontaneous population inversion emerges between the levels 1 and 2, which is maximal at $\mathcal T \simeq 0.063\, \gamma^{-1}$.
			Dashed lines indicate the equilibrium populations with respect to the cold temperature $T_{\text{c}}$, which would be approached in the long-time limit.
		\end{enumerate*}
		For all plots, we have set $\ell_1 = 1$ and $k_B T_{\text{c}} = \hbar\omega/2$.}
	\label{fig:2}
\end{figure*}

\subsection{Derivation} \label{subsec:derivation}

The mathematical derivation of our bound consists of the following steps.
First, we decompose the reservoir-induced ergotropy production into two components, $\ergoflux_t = \hbar\omega\gamma (\flux^1_t + \flux^2_t)$, and show that $\flux^1_t \leq 0$.
Second, by applying the rearrangement inequality to $\flux^2_t$, we isolate the contributions depending on the state of the system from those that are determined solely by the Hamiltonian of the working medium and the dissipation mechanism.
Finally, we cast the resulting expression into the form $\flux^2_t \leq \sum_{n=1}^N (r^n_t-r^{n+1}_t) \Phi_n$, thus proving Eq.~\eqref{eq:main_bound}.

In order to evaluate the thermal ergotropy production, we plug the master equation \eqref{eq:MD_MasterEquation} into Eq.~\eqref{eq:ThDyn_ErgotropyPower}.
Upon inserting factors of $\mathbbm 1 = \sum_{n=1}^N \ket{r^n_t} \bra{r^n_t}$ and using the relations $\comm{H_t}{L_t} = \hbar\omega L_t$ and $\comm{H_t}{L^\dagger_t} = -\hbar\omega L^\dagger_t$, we obtain the decomposition $\ergoflux_t = \hbar\omega\gamma (\flux^1_t + \flux^2_t)$ with
\begin{align}
	\flux^1_t &\equiv \nmsum \nu_t (r^m_t-r^n_t) (1+m-n) \abs{ \bra{r^n_t} L_t \ket{r^m_t} }^2
		\quad\text{and} \nonumber\\
	\flux^2_t &\equiv \nmsum r^n_t (n-m-1) \abs{ \bra{r^n_t} L_t \ket{r^m_t} }^2 .
\end{align}
Due to the ordering of the probabilities $r^n_t$, the expression $(r^m_t-r^n_t) (1+m-n)$ in the first term cannot be positive for any values of $n$ and $m$.
Since all other factors in this term are positive, it follows that $\flux^1_t \leq 0$.
It remains to analyze the second contribution $\flux^2_t$.
To this end, we introduce the partial sums $R^n_t \equiv \sum_{m=1}^n r^m_t$.
Using the inequality $r^n_t (n-m) \leq R^n_t - R^m_t$, we obtain
\begin{equation} \label{eq:NPC_BndIkin2}
	\flux^2_t \leq \nsum R^n_t \bra{r^n_t} \comm{L_t}{L^\dagger_t} \ket{r^n_t} - \nsum r^n_t \bra{r^n_t} L_t L^\dagger_t \ket{r^n_t} .
\end{equation}
In order to make this bound independent of the state $\rho_t$, we maximize the right-hand side of Eq.~\eqref{eq:NPC_BndIkin2} with respect to orthonormal vectors $\ket{r^n_t}$ in two steps~\cite{AllahverdyanEPL2004}.
First, we note that each term of the form $\bra\psi X \ket\psi$, with $X$ being a hermitian operator, is extremal as a function of the normalized vector $\ket\psi$ whenever this vector is an eigenvector of $X$.
The orthonormal sets of vectors $\ket{r^n_t}$ that maximize the two sums in Eq.~\eqref{eq:NPC_BndIkin2} must therefore form eigenbases of $\comm{L_t}{L^\dagger_t}$ and $L_t L^\dagger_t$, respectively.
For the second step, we apply the \emph{rearrangement inequality}~\cite{Hardy1952}, which fixes the ordering of these bases.
This procedure leads to
\begin{equation}
	\flux^2_t \leq \nsum \Bigl( R^n_t \pi_n\bigl( \bigl\{ \ell_{m-1}^2 {-} \ell_m^2 \bigr\} \bigr) - r^n_t \pi_n\bigl( \bigl\{ \ell_{m-1}^2 \bigr\} \bigr) \Bigr) ,
\end{equation}
where the function $\pi_n$ was defined after Eq.~\eqref{eq:phi_n}.
To bring this bound into the form of Eq.~\eqref{eq:main_bound}, we express the partial sums $R^n_t$ in terms of the probabilities $r^n_t$ and perform a summation by parts,
\begin{align}
	\flux^2_t &\leq \nsum \bigl( r^n_t-r^{n+1}_t \bigr) \Phi_n \quad \text{with} \\
	\Phi_n &= \ksum \Bigl( \sum_{\mathclap{j=k}}^N \pi_j\bigl( \bigl\{ \ell_{m-1}^2{-}\ell_m^2 \bigr\} \bigr) - \pi_k\bigl( \bigl\{ \ell_{m-1}^2 \bigr\} \bigr) \Bigr) . \nonumber
\end{align}
Using the relation $\sum_{j=1}^N \pi_j( \{ \ell_{m-1}^2{-}\ell_m^2 \} ) = 0$, we find that this expression for $\Phi_n$ is equivalent to Eq.~\eqref{eq:phi_n}.
The proof of the bound \eqref{eq:main_bound} is thus complete.

We note that, while this bound is effective to exclude systems with $\Phi_{\text{max}} \leq 0$ as potential working media, it does not imply a direct correspondence between the values of $\Phi_n$ and the amount of extracted work, see Fig.~\ref{fig:2}.
Furthermore, it is clear that, even for systems with $\Phi_{\text{min}} \equiv \min_n \Phi_n > 0$, it is still necessary to choose a suitable working protocol in order to achieve positive power output.
Therefore any sufficient condition for coherent power generation must inevitably depend on the applied driving protocols and can therefore not be as universal as our exclusion criterion.

\begin{figure*}
	\centering
	\includegraphics[scale=1]{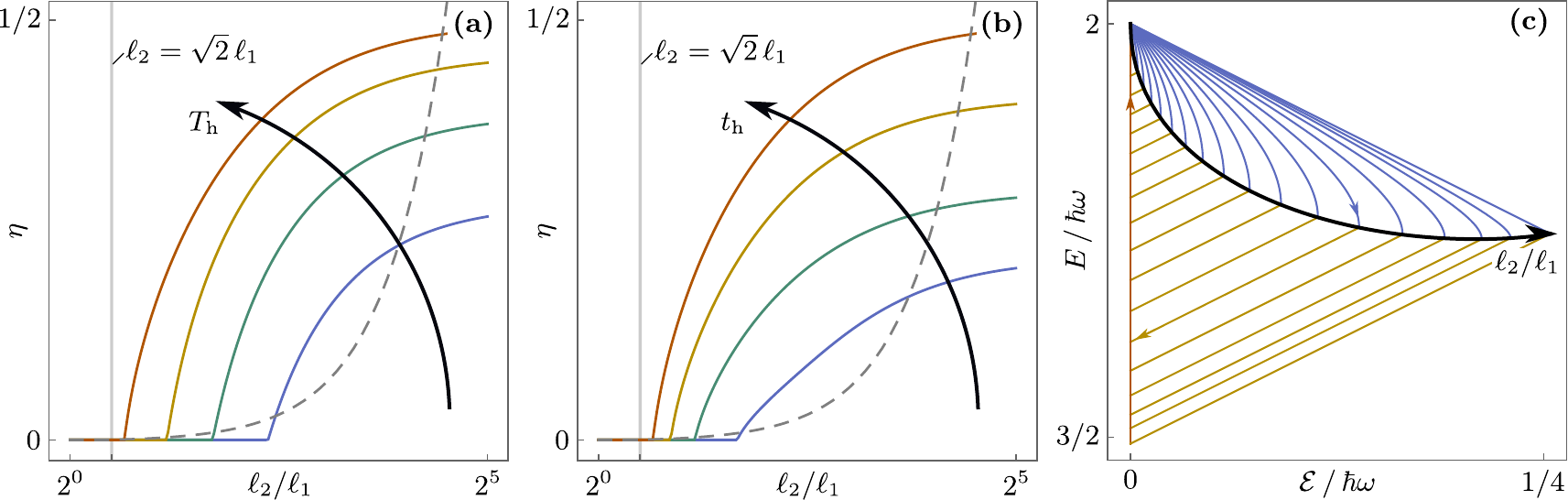}
	\caption[x]{Efficiency of the three-level amplifier.
		\begin{enumerate*}
		\item[(a) and (b)] Same plots as in Figs.~\ref{fig:2}a and \ref{fig:2}b, showing the efficiency $\eta$ instead of the work output $W$.
		\item[(c)] Ergotropy-energy diagram for various values of $\ell_2 / \ell_1$ between $\sqrt{2}$ and $2^{10}$.
			The thick arrow indicates the direction in which $\ell_2 / \ell_1$ increases.
			During one driving period at fixed $\ell_2 / \ell_1$, the cyclic state of the system follows the corresponding closed curve in clockwise direction, where the red (left), blue (top right) and yellow (bottom right) segments correspond to the first, second and third stroke, respectively.
			In every cycle, the extracted work $W$ is equal to the maximum of the ergotropy.
			The efficiency is therefore given by the ratio of the horizontal and the vertical extent of the curve.
			The plot shows the limit of infinite hot temperature $T_{\text{h}}$ and input time $t_{\text{h}}$, and we have set $\ell_1 = 1$ and $k_B T_{\text{c}} = \hbar\omega/2$.
		\end{enumerate*}}
	\label{fig:efficiency}
\end{figure*}

\section{Example} \label{sec:example}

\subsection{Three-Level Amplifier}

Having excluded qubits and harmonic oscillators as potential working media, we now turn to a three-level system, where $\Phi_{\text{max}}^{\text{3LS}} = \ell_1^2 - \ell_2^2$ if $\ell_2 \leq \ell_1$, $\Phi_{\text{max}}^{\text{3LS}} = \ell_2^2 - 2 \ell_1^2$ if $\ell_2 \geq \sqrt{2} \ell_1$, and $\Phi_{\text{max}}^{\text{3LS}} = 0$ otherwise.
Hence, three-level quantum ladders, whose jump weights satisfy
\begin{equation} \label{eq:NPC_3LS_bound}
	\ell_2 \leq \ell_1
	\quad\text{or}\quad
	\ell_2 \geq \sqrt{2} \ell_1
\end{equation}
are suitable candidates for cyclic quantum amplification.

To explore the physical picture behind the condition \eqref{eq:NPC_3LS_bound}, we now apply the protocol of Fig.~\ref{fig:1}c to a three-level system.
In the first stroke, the state~$\rho_t$ follows the master equation~\eqref{eq:MD_MasterEquation} with $T_t \equiv T_{\text{h}}$ for the time $t_{\text{h}}$. 
The temperature is then abruptly reduced to the cold level $T_{\text{c}} < T_{\text{h}}$.
The system relaxes at this temperature until its ergotropy becomes maximal, i.e., until the time $\mathcal T$, at which the difference $p^2_t - p^1_t$ of populations $p^n_t = \bra{n_t} \rho_t \ket{n_t}$ is maximal, see {Fig.~\ref{fig:2}c}.
At this time, the relaxation process is terminated and a $\pi$-pulse is applied, which swaps the populations of the lowest and the second level, thus generating the coherent work
\begin{equation} \label{eq:example_work}
	W = \ergo_{\mathcal T} = \hbar\omega (p^2_{\mathcal T} - p^1_{\mathcal T})
\end{equation}
and restoring the initial state of the system.

By applying this protocol repeatedly to an arbitrary initial state, we obtain the cyclic state of the system \cite{MenczelJPhysA2019,Note2}; the work output is then determined numerically.
The results of our analysis are summarized in Fig.~\ref{fig:2}, which reveals two key effects.
First, beyond a certain threshold value, which depends on $T_{\text{h}}$ and $t_{\text{h}}$, the average work $W$ grows monotonically as a function of the ratio~$\ell_2 / \ell_1$.
This behavior arises from an increasing separation between the characteristic relaxation times $\tau_2 = 1 / (\gamma \ell_2)$ and $\tau_1 = (1 / \gamma \ell_1)$ of the upper and the lower level.
If $\tau_2 \ll \tau_1$, the population of level 3 can be essentially transferred to level 2 before level 1 is significantly affected.
In this regime, a pronounced population inversion emerges, leading to a large output $W$.
Second, we find that an increasing amount of work can be extracted if either the hot temperature $T_{\text{h}}$ is raised or if the duration $t_{\text{h}}$ of the input stroke is extended;
	at the same time, the threshold value of $\ell_2 / \ell_1$ decreases.
This phenomenon can be understood by observing that the level populations after stroke 1 become more homogeneous for larger values of $T_{\text{h}}$ and $t_{\text{h}}$.
Hence, less population has to be redistributed during the second stroke to create a strong inversion.
In the limiting case $T_{\text{h}}, t_{\text{h}} \rightarrow \infty$, all three levels are equally populated after the first stroke and the bound~\eqref{eq:NPC_3LS_bound} becomes tight.

\subsection{Efficiency}

So far we have studied the work output of our amplifier without accounting for its thermodynamic cost.
For heat engines, this cost corresponds to the heat input $Q_{\text{h}}$ supplied by the hot reservoir \cite{ScovilPhysRevLett1959, BrandnerPhysRevE2016}.
In our example, it is therefore given by the increase in the system energy $E_t$ during the first stroke.
Note that the energy required to create the $\pi$-pulse in the third stroke should not be counted as input, since it remains in the coherent light field together with the extracted work.
The thermodynamic efficiency of our cyclic quantum amplifier is thus given by the ratio
\begin{equation}
	\eta \equiv W / Q_{\text{h}} .
\end{equation}
It satisfies the Carnot bound
\begin{equation} \label{eq:carnot}
	\eta \leq \eta_{\text{C}} \equiv 1 - T_{\text{c}} / T_{\text{h}}
\end{equation}
due to the second law of thermodynamics
\begin{equation}
	\Delta S = -\frac{1}{T_{\text{h}}} Q_{\text{h}} - \frac{1}{T_{\text{c}}} Q_{\text{c}} \geq 0 ,
\end{equation}
where $-Q_{\text{c}}$ is the heat transferred to the cold reservoir and $\Delta S$ the total entropy production in one period.

Our results in {Fig.~\ref{fig:efficiency}} show that the efficiency behaves qualitatively similar to the work output discussed before.
Specifically, it increases monotonically as a function of the ratio $\ell_2 / \ell_1$, the hot temperature $T_{\text{h}}$ and the relaxation time $t_{\text{h}}$.
In the limit $\ell_2 / \ell_1, T_{\text{h}}, t_{\text{h}} \to \infty$, it reaches the upper bound
\begin{equation} \label{eq:eff_bound}
	\eta \leq \eta_{\text{max}} \equiv 1 / 2 ,
\end{equation}
which is smaller than the Carnot bound \eqref{eq:carnot} for the temperatures used in the plots of Fig.~\ref{fig:efficiency}.
This constraint arises because our three-level amplifier creates ergotropy in the second stroke by transferring population from the third to the second energy level.
Due to the equidistant level spacing, this process is accompanied by the loss of an equal amount of internal energy, that is,
\begin{equation} \label{eq:eff_bound_derivation}
	\Delta_2 \mathcal E \leq -\Delta_2 E ,
\end{equation}
as can be seen clearly in {Fig.~\ref{fig:efficiency}c}.
Here, $\Delta_2 X$ denotes the change of a quantity $X$ during the second stroke.
To formally derive the inequality \eqref{eq:eff_bound_derivation}, it suffices to verify that either $\mathcal J_t \leq 0$ or $\mathcal J_t + \Tr{\dot\rho_t H_t} \leq 0$ holds for any diagonal state of the system.
We stress that the efficiency bound (23) is specific for three-level systems.
Cyclic quantum amplifiers with more energy levels can reach efficiencies larger than $1/2$.

\section{Concluding Perspectives} \label{sec:outlook}

\begin{figure}
	\centering
	\includegraphics[scale=1]{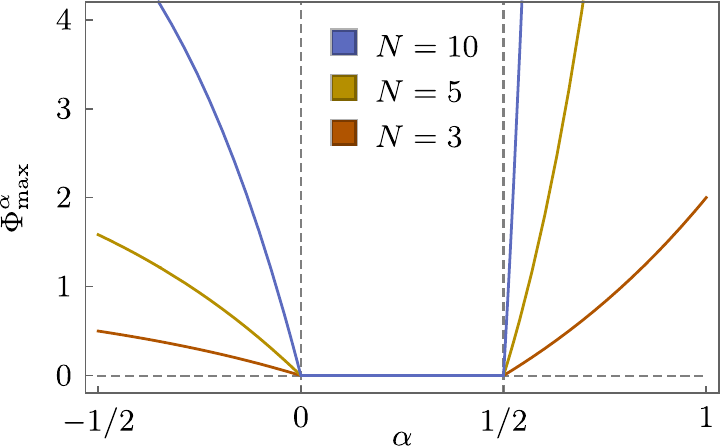}
	\caption{Working criterion~\eqref{eq:NPC_BoundPhiMax} for algebraically scaling jump weights $\ell_n = n^\alpha$ and different numbers $N$ of energy levels.
		Coherent power generation can be achieved only for $\Phi_{\text{max}} > 0$.
		The dashed lines are guides to the eye.}
	\label{fig:3}
\end{figure}

Having studied the three-level system in detail, we now consider a more general type of quantum ladder, for which $N \geq 3$ is arbitrary and the weights $\ell_n$ depend algebraically on the level index, i.e., $\ell_n = n^\alpha$.
The corresponding coefficients $\Phi^\alpha_{\text{max}}$ are plotted in Fig.~\ref{fig:3} for different numbers of energy levels.
Notably, we find that, irrespective of $N$,
\begin{equation} \label{eq:NPC_BoundAlpha}
	\Phi^\alpha_{\text{max}} = 0
		\quad\text{for}\quad 
	0 \leq \alpha \leq 1/2 .
\end{equation}
This result suggest that the squared weights $\ell_n^2$ must either decrease with $n$ or increase at least linearly to enable the spontaneous creation of population inversion.
Whether or not this observation can be corroborated for more complicated relations between $\ell_n$ and $n$ remains as an open question.

Turning to more general situations, we note that the versatile technique that we have developed to derive our working criterion~\eqref{eq:NPC_BoundPhiMax} can be easily adapted for setups with multiple reservoirs or composite working systems that consist of a collection of non-interacting quantum ladders.
Further extensions of our scheme might even make it possible to consider non-equilibrium reservoirs~\cite{ScullyScience2003, RossnagelPhysRevLett2014, ManzanoPhysRevE2016, KlaersPhysRevX2017, NiedenzuNatCommun2018} or strongly driven systems, for which the master equation~\eqref{eq:MD_MasterEquation} has to be replaced by a Floquet-Lindblad equation~\cite{BreuerPhysRevA1997, AlickiPhysRevA2006, KosloffEntropy2013}.
In principle, the qualitative behavior observed in our case study can be expected to persist for more general systems;
	that is, a strong separation of relaxation time scales and the preparation of the working system in a state with nearly flat level populations during the input stroke should generically improve the performance of cyclic quantum amplifiers.
As the technology that is available to realize engineered quantum systems with fine-tuned interactions is improving rapidly \cite{PekolaNatPhys2015, MillenNewJPhys2016}, we expect that these assertions will become accessible for experimental investigation in the near future.

These possibilities show that our work provides both a new approach to assess the viability of coherence-based heat engines and a valuable starting point for future investigations seeking to further explore the mechanisms of thermal energy conversion in the quantum regime.
Our results thereby corroborate the emerging picture that coherent power generation is a technically demanding process, which requires a well-tailored setup.
In fact, it was shown only recently that thermodynamic cycles cannot produce coherent work in the limits of linear~\cite{BrandnerPhysRevLett2017} and adiabatic~\cite{BrandnerPhysRevLett2020} response, i.e., if either the overall amplitude or the frequency of the driving field and temperature variations is small. 
Here, we have taken a first step towards a more complete characterization of the necessary working conditions for periodic thermal devices that deliver coherent energy output.
In particular, our working criterion~\eqref{eq:NPC_BoundPhiMax} shows that, even far from equilibrium, cyclic quantum amplifiers are subject to much stronger restrictions than conventional cyclic heat engines.

\begin{acknowledgments}
K.B. acknowledges support from the Academy of Finland (Contract No.~296073).
This work was supported by the Academy of Finland (Projects No.~308515 and No.~312299).
All authors are associated with the Centre for Quantum Engineering at Aalto University.
\end{acknowledgments}

\end{document}